# Spacecraft Coatings Optimizing LiDAR Debris Tracking and Light Pollution Impacts

**Julia Hudson***, **Eric Jones****

* New Trier Township High School
** Department of Physics and Astronomy, Stony Brook University

*Abstract*- Space safety and astronomy are at odds. The problem posed by space debris and derelict satellites in the low Earth orbit is an existential threat to all space operations. These dangerous objects in space are more easily tracked with ground-based LiDAR if they are highly reflective, especially in the near-infrared (NIR) range. At the same time, reflective objects in orbit are the bane of ground-based astronomers, causing light pollution and marring images with bright streaks. How can this tension be resolved? The hypothesis tested is that a near-infrared-transparent (NIRT) coating which is opaque in the visible light range and transparent in the NIR range is a promising candidate for use in satellite construction. This experiment tests whether typical spacecraft surfaces such as anodized aluminum or multi-layer insulation (MLI) with a NIRT coating applied will absorb visible light and reflect NIR. The findings confirm the efficacy of the NIRT coating for this purpose, reducing visible light reflection by 47% (+/-3%) and increasing reflection in the NIR by 7% (+/-2%). This promising novel NIRT coating may help provide a path forward to resolve the tension between astronomy and the space industry.

*Index Terms*- Space debris; LiDAR; spacecraft coatings; NIR transparent; laser ranging; Kessler Syndrome
*Contact*- Julia Hudson (julia.hudson@outlook.com)

## I. INTRODUCTION

The European Space Agency estimates that there are currently over 130 million pieces of space debris of at least 1 mm in diameter in the low Earth orbit, the region of space from 400 to 2,000 km in altitude (ESA 2023a). With typical speeds of about 8,000 m/s, even a fragment as small as a bolt can cause catastrophic damage to spacecraft in orbit. Being able to track the position and velocity of these fragments in orbit is the key to be able to avoid this debris, or even eventually clean it up. While radar and optical sighting have some role in locating debris, light detection and ranging (LiDAR) is the most promising, having proven the capability to track thousands of objects with an accuracy of 1 cm or less (Degnan 1993).

The danger of inaction on space debris is extreme. As early as 1978, Dr. Kessler at NASA warned that the accumulation of space debris and derelict satellites in the low Earth orbit could eventually cause a runaway cascade of collisions, destroying most orbital assets and shutting down space operations for centuries (Kessler 1978). This catastrophic scenario became known as the "Kessler Syndrome." Since then, space agencies and industries have worked to counter the threat posed by space debris.

One key strategy for mitigating this danger is to expand the global capacity to detect and track space debris in orbit with ground-based LiDAR stations. With sufficient power, these long range lasers can track objects in the low Earth orbit, and the use of short wavelengths in the micrometer range means that even small debris fragments can be resolved (Mehrholz *et al.* 2002).

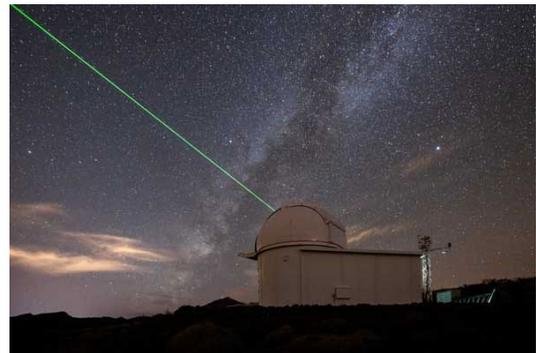

*Figure 1. ESA's Izaña-1 laser ranging station (ESA 2023b)*

There currently exists a network of ground-based LiDAR stations known as the Laser Ranging Service (ILRS), which provides laser ranging data to support research in geodesy, geophysics, and other scientific and operational uses (ILRS 2016). The ILRS network has been used to track satellites in the past, and the expectation is that it will be more active in space debris ranging in the future.

Starting with the original LiDAR installation at NASA's Goddard Space Flight Center, the primary method for tracking objects in orbit has been to aim the beam at multi-faceted mirrors called retroreflectors attached to satellites, ensuring that the signal from the ground-based laser was returned. However, if a spacecraft has been reduced to debris from a collision or breakup, then no retroreflector will be present, and the ranging will rely on the reflectivity of the fragment on its own. This is known as "uncooperative" debris.

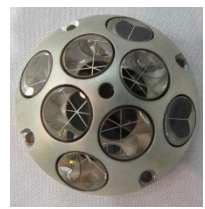 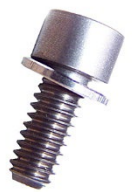

*Figure 2: A typical example of a spacecraft retroreflector (left), from the NASA Lunar Reconnaissance Orbiter (Leonard 2019), and an illustrative example of potential future "uncooperative" debris (right), a spacecraft bolt from Reid Products (Reid 2023)*

Being able to track uncooperative debris then is the next challenge for the space industry. Numerous groups have proven that the existing ILRS network, with sufficient planning and coordination, can accomplish this task, as long as the debris is sufficiently reflective (Lejba *et al.* 2018).

Therefore, from the perspective of space operators, the obvious solution would be to make the entire space fleet "shiny" in order to increase the reflectivity the whole or its parts. However, this runs head-on into the goals of Earth-based astronomers, who seek to keep the night sky dark. Depending on the geometry of the sun on a given night, reflective satellites (or fragments) will visibly glow, disrupting hard-sought observations.

The millions of reflective objects in orbit are already causing headaches for astronomers. The reflection of sunlight from moving objects creates point sources which often appear as "streaks" across images taken of the night sky. These streaks are most prevalent during the first and last hours of the evening, and for wide-field exposures on a large telescope, can compromise 30-40% of the images taken (Hainaut 2020).

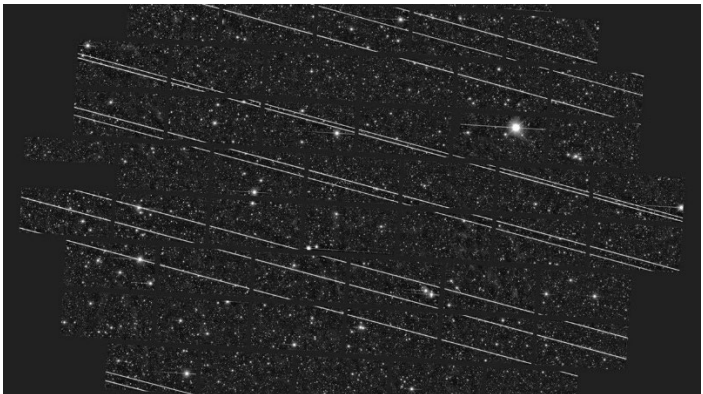

*Figure 3. Astronomers at the Cerro Tololo Inter-American Observatory captured SpaceX's Starlink satellites streaking across the night sky in November. Nineteen streaks are visible in the 333-second exposure. Credit: National Optical-Infrared Astronomy Research Lab (Aerospace 2020)*

Beyond the point sources, the aggregate effect of millions of reflective objects in orbit is to contribute to the overall night sky brightness, sometimes called "skyglow." It is estimated that the impact of these objects currently is causing a zenith visual luminance of 20 μcd m−2, which is a 10% increase above the luminance of the natural night sky (Kocifaj *et al.* 2021). This threshold is considered to be a light pollution 'red line' requiring concerted action to address, if astronomy is to remain viable.

How can a compromise be found? How could materials be employed that are good LiDAR targets but won't bounce sunlight? One solution would be to imagine a coating that absorbed visible light, but was reflective, or at least transparent, in the NIR range. This might check the box for both factions.

Such materials do exist. There have been pigments developed that fit the bill exactly, transparent to NIR and appearing black in visible light, with a limited number of these NIRT coatings being sold commercially. These products generally target terrestrial applications, such as dark-colored paint for a car that will be LiDAR-reflective, or signage that is only visible in the NIR. There is also ongoing work on NIRT materials based on various organic and inorganic chemistries, as well as novel molecular compounds such as diketopyrrolopyrrole pigments (Yu *et al.* 2023). Despite the ongoing research in developing LiDAR-optimized coatings for automobiles and new materials to create pigments, we were not able to discover any research regarding space-borne applications for NIRTs.

For this experiment, we sought out a commercial product in order to have consistent and testable quantities to work with. The subject NIRT chosen for this experiment was a product branded "Magic Black Paint," produced by Anytime Sign, Inc (also doing business as Infrared Coatings, Inc.). The chemistry underlying its NIR-transparency is proprietary and could not be disclosed in a public paper. This Magic Black NIRT coating is as sold as an aerosol spray paint or prefabricated in mylar sheeting. As can be seen in the specification materials from the vendor in *Figure 4*, the pigment is highly transparent at NIR wavelengths.

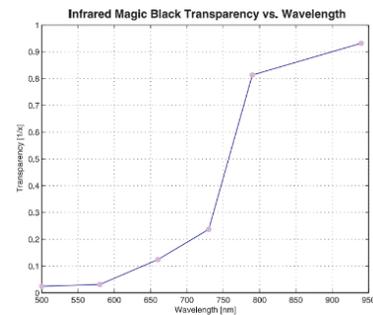

*Figure 4: Transparency vs light frequency for the Magic Black Coating, with frequency of typical LiDAR laser noted by author. Note that the material is 20% or less transparent in the visible light range and over 90% transparent in the NIR range used for LiDAR (1060 nm). Source: Anytime Sign, Inc. (Infrared 2023)*

Returning to the type of laser used for this experiment, there are a number of types of lasers employed by the ILRS for these different purposes. The most common wavelength used at ground-based LiDAR stations for satellite laser ranging (SLR) is 532 nm (frequency-doubled from 1064 nm), though the six stations actively involved in debris laser ranging (DLR) use the un-doubled 1064 nm in the NIR range. This wavelength is considered the most promising for tracking space debris, due to superior laser power, atmospheric transmission, and ability to range targets with weak signals, even in the daytime (Meyer 2022). Therefore, use of the 1064 nm is what will be demonstrated in this experiment.

The pulse repetition frequency (PRF) for the laser used in SLR and DLR is also an important consideration. A derivation beginning with the radar link equation demonstrates that increasing the PRF of a laser used for ranging will increase its sensitivity (Long *et al.* 2022). Very high PRFs, with pulse lengths as low as the picosecond level, have been tested for DLR applications (Zhang *et al.* 2021). Given that successful space debris measurements have been accomplished with a 1 kHz repetition rate (Kirchner *et al.* 2012), the 10 ns pulse length and 100 MHz PRF used in this experiment will represent a state-of-the-art configuration for SLR and DLR stations.

The hypothesis tested in this paper will be whether NIRT materials would indeed be candidates for satellite and debris tracking applications. This will require coated components to demonstrate high reflectivity of a proxy LiDAR laser and low reflectivity in the visible light spectrum.

Finding a solution to increase the visibility of space debris, while not destroying astronomy in the process, is critically important as the amount of space junk is forecast to soar in decades to come. Whereas the population of active satellites was about 2,000 in 2020, just a single operator, Starlink, has permits to launch 42,000 satellites over the next decade (Ma and Zhang 2022). Each of these satellites in the Starlink or other planned "constellations" will eventually be a source of future debris. Will the world be ready when the amount of space debris, doubles, triples, or increases ten-fold in the years to come? How can the space industry and the scientific endeavor of astronomy co-exist in the era of massive space commercialization?

## II. METHODS

Essentially, two experiments were conducted: The first was a test of the reflectivity of targets, made from indicative spacecraft materials, illuminated by visible light. The second experiment was to test the reflectivity of the same targets again with an Yb:fiber frequency comb laser (approximating the specifications of LiDAR tracking stations). Eight different types of targets were constructed to represent a range of typical spacecraft materials. The hypothesis was tested by comparing the outcomes from uncoated targets and targets coated with a NIRT material.

### A. Visible light setup

A tungsten lamp was used to provide a broad-spectrum light source that would effectively simulate sunlight across the visible spectrum and infrared spectrum. The light was collimated using a pair of apertures and a plano-convex lens. Once the light was collimated, it was reflected off sample and into the VIS/NIR spectrometer. The setup is shown in *Figure 5*:

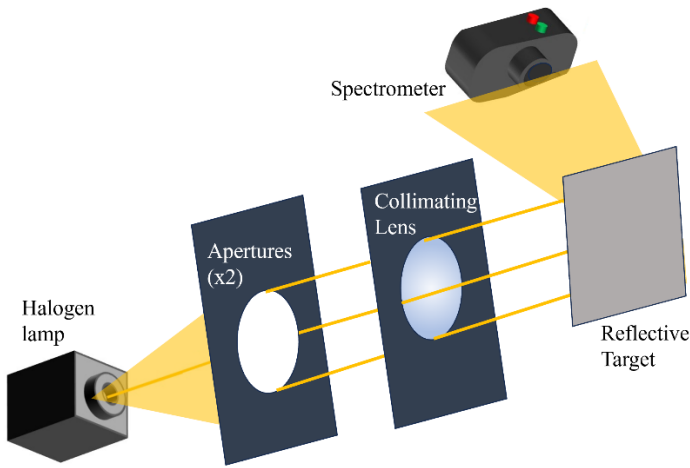

*Figure 5. Schematic of visible light collimation and reflectivity measurement. From left to right is shown the tungsten lamp, apertures, collimation lens, and sample. Light reflected from the target sample is directed toward a visible light spectrometer.*

### B. Visible light reflectivity (intensity) measurement

Measurement of the visible light was done with a Broadcom AFBR-S20M2VN spectrometer, tracking the reflectivity across wavelengths in its range (400-1100 nm) produced by the halogen lamp. The intensity at each wavelength was recorded, which could then be compared on a relative basis between targets.

### C. NIR laser setup

A tabletop laser was constructed to provide a beam at 1064 nm that would be representative of a ground-based laser ranging station used for DLR. The design used for this experiment was a custom-built high-power low noise Yb:fiber frequency comb laser with a repetition in the 100 MHz range, as described by Dr. Yuning Chen in his thesis (Chen 2018).

The fiber section provides gain and nonlinearity while the components in the free space section compensate the dispersion of the fiber, manipulate the polarization, and actuate on the pulse's round-trip group delay and carrier-envelope offset phase. The design is diagrammed in *Figure 6*:

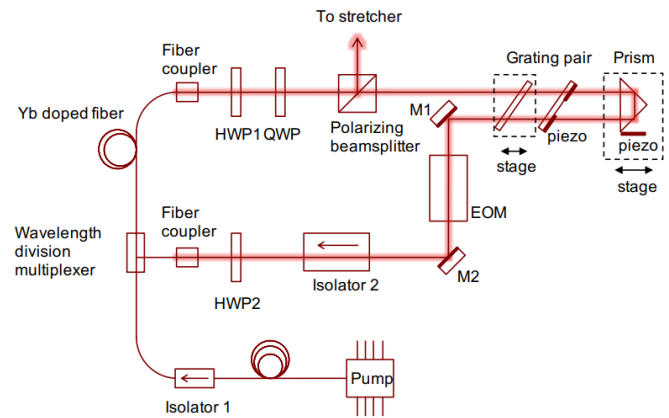

*Figure 6. The block diagram shows the optical layout of the Yb:fiber oscillator. EOM: electro-optic modulator, G: grating, M: mirror, HWP: half. (Chen 2018)*

Dr. Chen's design provides for a mode-locked spectrum of the oscillator with a peak spectrum in the range of the 1064 nm target:

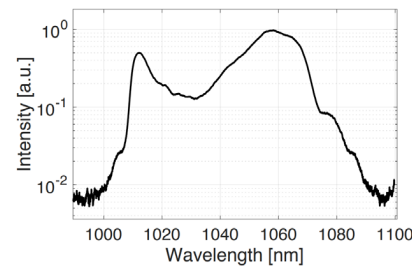

*Figure 7. Yb:fiber oscillator spectrum (Chen 2018)*

The beam was measured with a digital phosphor oscilloscope, showing the correct frequency of pulses with ~100 MHz frequency (~10 ns period).

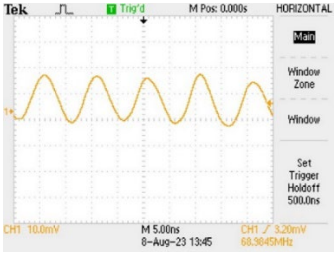

*Figure 8. Oscilloscope measurement of Yb:fiber laser*

Therefore, this spectrum, frequency, and pulse length produced by our tabletop laser will be a good proxy for the actual laser at a ground-based station used for DLR.

D. <u>NIR light reflectivity (intensity) measurements</u>

The beam from the Yb-fiber oscillator is passed through the output coupler and directed towards the reflective target via a set of mirrors. A Coherent FieldMate laser power meter is positioned to receive the beam after it reflects from the target material. The diagram in *Figure 9* describes the setup.
.

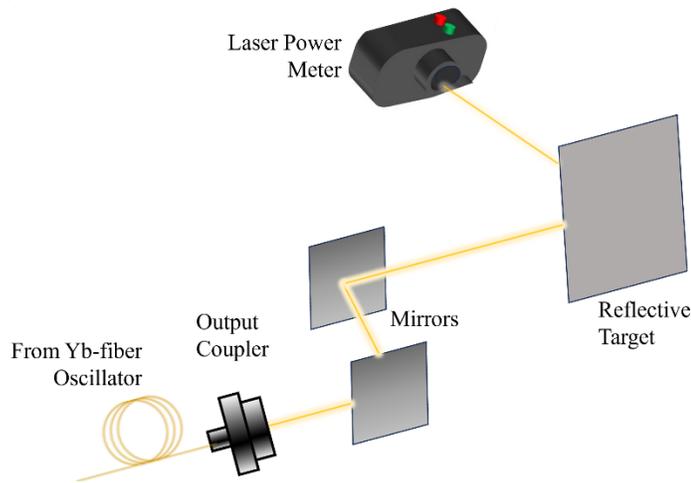

*Figure 9. Schematic of NIR laser reflectivity measurement. From left to right is shown the laser light source, guiding mirrors, and sample. Light reflected from the target sample is directed towards a laser power meter.*

The detector was tuned to measure light at the 1064 nm wavelength, and report an intensity count, which will be compared between targets on a relative basis.

E. <u>Selection of targets</u>

Targets for the reflectivity test were chosen in order to simulate a reasonable selection of typical spacecraft materials. The exterior surfaces of a spacecraft could be any of these materials, though often multi-layer insulation (MLI) materials cover much of the surface to meet thermal requirements. Aluminum in various alloys is widely used for spacecraft construction and is expected to represent most of the mass of fragments in the event of a satellite breakup or explosion (Cowardin *et al.* 2021). *Table 1* summarizes the target types constructed:

*Table 1. List of targets materials tested*

| Base Material | Notes |
|---|---|
| Aluminum. Alloys 2024, 6061, and 7075 were tested | The most common alloys used in spacecraft construction |
| Target Type (Surface over Base Material) | Notes |
| None | Bare aluminum surface |
| Beta cloth | Silica fiber fabric with low absorptivity, less reflective than aluminum |
| Black MLI | Multi-layer insulation (MLI) coverings are also known as a "blankets." Sample materials were plastic Kapton films with aluminum or copper coating vapor deposited on the surface, either black or metallic side outwards. MLIs are sometimes combined with Beta Cloth in applications. |
| Black MLI and Beta Cloth | |
| Silver MLI | |
| Silver MLI and Beta Cloth | |
| Copper MLI | |
| Copper MLI and Beta Cloth | |

Therefore there were a total of eight types of targets tested, with each target type being tested with all three alloys of aluminum, and also with and without the NIRT coating, meaning 48 total tests were done with each light source.

The anodized aluminum alloy sheets were procured from McMaster-Carr. The MLI and beta cloth samples were provided by the Aerospace Fabrication company. The NIRT material was provided in the form of Magic Black spray pain and Magic Black adhesive-backed mylar procured from Anytime Signs, Inc.

F. <u>Preparation of targets</u>

The bulk sheets of the aluminum alloys procured were cut into 3cm x 3cm squares using a band saw. The MLI and beta cloth samples were also cut with scissors to the same size as the metal targets.

For the control set, intended to represent actual spacecraft surfaces without the NIRT coating, unmodified aluminum targets were used, as well as the same aluminum targets with black MLI, silver MLI, copper MLI, and beta cloth attached. A small amount of conventional super glue was used in the targets' corners to fix the MLI and beta cloth in place.

For the experiment set, all of the same targets were constructed as the control set, and then all of these targets were spray-painted with Magic Black. The spraying was performed on a consistent basis, with a 2-second burst at a six-inch distance.

The targets were held in place with a customized 3-d printed mount, that was fixed in place. This mount allowed for the targets to swapped out in succession without changing the angle of incidence or distance to the light source.

### G. Collection of measurements

All of the measurements for each light source, the visible light and the NIR laser, were all done in a single session with no change to the experimental setup. The laboratory was in a windowless basement room, and ambient light was unchanged, so there should be no meaningful variation in background light. For the spectrometer measurements (visible light), the intensity measurements were collected in 2000 wavelength bins ranging from 455 nm to 1105 nm. For the laser power meter measurements (NIR light) a single intensity reading was taken. Error bars were estimated to be +-10% due to potential variations in exactly how the materials were placed in the target mount, as well as potential variability in the halogen light or NIR laser intensity over time.

The combination of 8 target types, three alloys for each (see *II-E*), and two coatings (no coating and spray) results in 48 measurements for each of the light sources (halogen and laser). For both the visible and NIR light data, the results from the three aluminum alloys were added together for an aggregate result.

## III. RESULTS

### A. Visible light reflectivity

Data for the 48 measurement combinations described in section *II-G* was collected using the halogen visible light source and spectrometer. The results are plotted in *Figure 10* for intensity vs. wavelength for one of the aluminum alloys, 6061, with the reduction in reflectivity from the NIRT coating highlighted. The same data was collected for each of the other alloys.

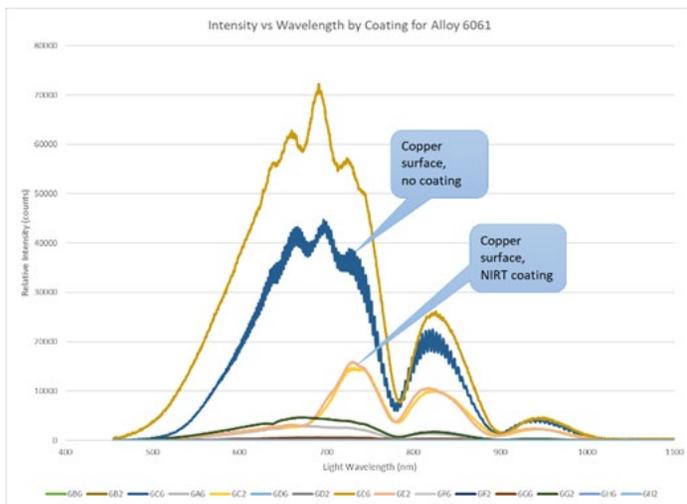

*Figure 10 Reflectivity of targets to visible light by wavelength, surface, and coating*

To determine an aggregate reflectivity for each combination of alloy, surface, and coating the area under the spectroscopy curve (from 380-740 nm) was determined, providing for a single reflectivity data point. This is intended to match with the overall goal of the research, which is to determine means of reducing overall visible light pollution from satellites in orbit. *Figure 11* shows an indicative sampling of the measured intensity of visible light for a few combinations, with the percentages representing the change in intensity between the sample with and without the NIRT coating.

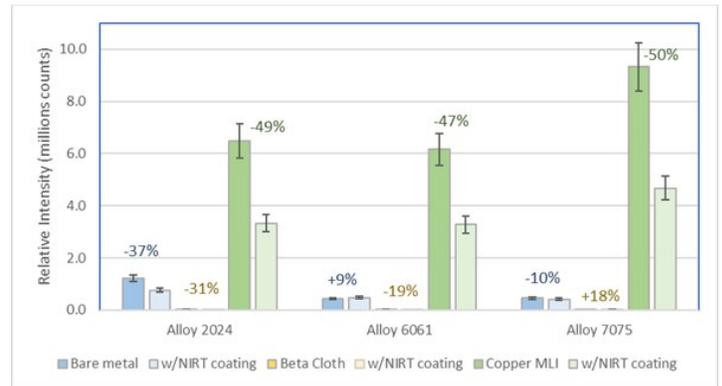

*Figure 11. Reflectivity of targets to visible light with and without NIRT coating, including percentage change of reflectivity from the coating*

For the summarized data, the intensity counts of the three aluminum alloys were added together to provide for an effective average basis.

### B. NIR light reflectivity

The same 48 target combinations of surface and coating were measured using the NIR laser and detector. *Figure 12* shows an indicative sampling of the measured intensity of NIR light for a few combinations, with the percentages representing the change in intensity between the sample with and without the NIRT coating.

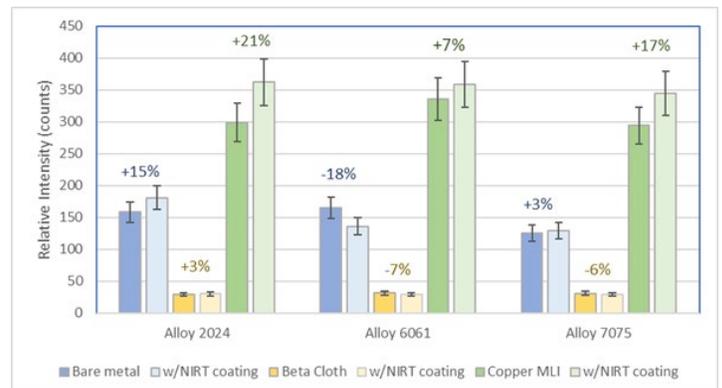

*Figure 12. Reflectivity of targets to NIR light with and without NIRT coating, including percentage change of reflectivity from the coating*

As with the visible light data, for the summarized NIR data, the intensity counts of the three aluminum alloys were added together to provide for an effective average basis.

### C. Aggregated data

A methodology was developed to provide an overall "score" for how well the NIRT material performed towards the two goals of: (1) reducing visible light reflectivity and (2) maximizing NIR reflectivity.

For the visible light, the relative reflectivity spectroscopy data from section *III-A* was summarized by calculating the area under the curve from 380 nm to 760 nm (the visible spectrum). A percentage benefit for the NIRT material was calculated, defined as:

$$\frac{(\text{intensity without coating}) - (\text{intensity with coating})}{\text{intensity without coating}}$$

[noting that a reduction in intensity from the coating will provide for a positive "score."]

For the NIR laser light, the intensity measurements were used as is without any adjustments. A percentage benefit for the NIRT material was calculated, defined as:

$$\frac{(\text{intensity with coating}) - (\text{intensity without coating})}{\text{intensity without coating}}$$

The signs are reversed between the two equations, as in the first case reductions are sought and in the second case increases are sought. Note also that the data for the three alloys of anodized aluminum (2024, 6061, and 7075) were very similar, and were simply summed in this analysis.

D. Comparison of NIRT to uncoated surfaces

i. Efficacy at reduction visible light reflectivity

The aggregated data (sum under spectrum curve) was compared in the uncoated and NIRT-coated material cases across all of the surface types, with the calculated percentage benefit:

*Table 2. Efficacy of NIRT Coating to Reduce Reflection of Visible Light*

| | Reflectivity of Visible Light (counts) | | |
|---|---|---|---|
| | No coating | NIRT | NIRT benefit |
| **Bare Aluminum** | 5.7E+06 | 3.7E+06 | 36% (+/-2%) |
| **Beta Cloth** | 8.5E+04 | 5.4E+04 | 37% (+/-2%) |
| **Black MLI** | 1.8E+06 | 5.7E+06 | -214% (+/-12%) |
| **Black MLI, Beta** | 7.0E+04 | 5.9E+04 | 15% (+/-1%) |
| **Silver MLI** | 1.1E+08 | 1.8E+07 | 84% (+/-5%) |
| **Silver and Beta** | 8.8E+04 | 6.0E+04 | 32% (+/-2%) |
| **Copper MLI** | 6.3E+07 | 1.1E+07 | 83% (+/-5%) |
| **Copper and Beta** | 9.8E+04 | 5.8E+04 | 41% (+/-2%) |
| **AVERAGE COATING BENEFIT (ex outlier)** | | | 47% (+/-3%) |

This particular combination of Black MLI coated with NIRT proved to be unusually reflective, and will be dropped from consideration for candidate materials (and excluded from this analysis).

The data show that the NIRT material is effective at reducing the reflection of visible light from most spacecraft surfaces, with the highest benefit to copper and silver MLI surfaces, and the lowest benefit to a combination black MLI beta cloth surface, and negative benefit in the black MLI case.

ii. Efficacy at maximizing NIR reflectivity

The data was compared in the uncoated and NIRT material cases across all of the surface types, with the calculated percentage benefit:

*Table 3 Efficacy of NIRT Coating to Improve Reflection of Near-Infrared Light*

| | Reflectivity of NIR Light (counts) | | |
|---|---|---|---|
| | No coating | NIRT | NIRT benefit |
| **Bare Aluminum** | 450 | 450 | 0% (+/-6%) |
| **Beta Cloth** | 84 | 86 | 2% (+/-6%) |
| **Black MLI** | 930 | 1,100 | 18% (+/-6%) |
| **Black MLI, Beta** | 100 | 120 | 20% (+/-6%) |
| **Silver MLI** | 100 | 110 | 10% (+/-6%) |
| **Silver and Beta** | 1,100 | 1,100 | 0% (+/-6%) |
| **Copper** | 91 | 97 | 7% (+/-6%) |
| **Copper and Beta** | 88 | 98 | 11% (+/-6%) |
| **AVERAGE COATING BENEFIT (ex outlier)** | | | 7% (+/-2%) |

The data show that the NIRT material is effective at not impairing, and most cases improving, the reflection of NIR light from most spacecraft surfaces, with the highest benefit to copper MLI and copper & beta cloth, and no benefit (though no loss either) with the bare aluminum case.

IV. CONCLUSION

The experiment determined that the Magic Black Spray Paint as a NIRT coating met the objectives for (1) reducing reflectivity to visible light and (2) increasing reflectivity for the near-infrared laser:

**(1) Reduction in reflection of visible light (380-740 nm): 47% (+/-3%)**

**(2) Increase in the reflection of NIR (1064 nm) laser: 7% (+- 2%)**

This means that if this coating could be commercialized for spacecraft use, it could be a solution that worked for both astronomers and for those interested in tracking intact craft and debris.

## V. DISCUSSION

The findings are positive, that the NIRT coating significantly reduced the reflection of visible light from all types of spacecraft materials (except for black MLI). The increased reflectivity of NIR light was a surprising result, which speaks further to the efficacy of the NIRT coating to facilitate LiDAR tracking. At least based on these initial findings, objects in orbit could be made darker, but still trackable, with widespread use of NIRT coatings. However, a number of operational challenges still need to be addressed.

One key consideration for the use of NIRT materials as spacecraft coatings is that it would need to be determined if the new coating negatively impacted any of the operating parameters, such as thermal, electrical, or mechanical. An example of this concern occurred with the introduction of SpaceX's DarkSat program. In response to the same concerns from astronomers referenced in the introduction, a number Starlink satellites were coated with a dark-colored material, intended to be highly nonreflective. While the coating did succeed in lowering the light pollution produced by the satellites, it resulted some heating problems due to increased radiation absorption.

We expect that the NIRT material proposed here would not have the same problem, as they are transparent in the NIR and still somewhat reflective in the visible. Therefore we would anticipate excess heating to be far less of a problem, though that assumption would need to be tested further before large-scale deployment.

Also, there may be a distinction between using the NIRT coating on metal surfaces such as anodized aluminum and other locations. As noted earlier, certain spacecraft surfaces are covered with MLI and beta cloth materials, sometimes called "blankets." These blankets are placed over key areas for specific mission requirements. It would need to be determined if the NIRT coating interfered with the useful properties of those materials.

How the NIRT coating would be applied in practice to spacecraft is also an open question. While this experiment used a spray paint product to deliver the coating, this would not be suitable for satellites, since paint can chip and fleck, resulting in space debris hazards. We would expect that for orbital use of these materials, a deposition or similar process would be used to fuse the materials to surfaces, similar to what is done with other coatings. Alternately, the company, called Infrared Coatings, that produces the Magic Black spray, also produces coated mylar sheets with the same properties. This could also potentially be an option, or perhaps could be incorporated as a layer in MLI materials.

In any case, it is expected that engineering solutions could be found to facilitate the use of NIRT on spacecraft. And given the strong promise shown by these materials to make spacecraft and space debris more visible for the LiDAR tracking network, while not compromising astronomy on Earth, it seems to be an avenue worthy of progressing in the near term. These novel NIRT coatings may hopefully become a useful tool to ensure that space safety and astronomy need not be at odds.

## ACKNOWLEDGMENT

Thank you to Dr. Eric Jones, Dr. Harold Metcalf, Dr. Marty G. Cohen, Mike Wahl, and the Laser Teaching Center for constant guidance and support. Thank you to Rudy Popper and the Allison Research Group for helping me use the femtosecond laser and thank you to Julia Codere and Michael Belmonte from Tom Weinacht's Ultrafast group for helping me with the visible spectrometer. Thank you to Brent Anderson from Aerospace Fabrication and to Patrick McIntyre from Anytime Sign for generously sending me free samples and inspiring my research. Thank you to the Simons Summer Research Program and Stony Brook University for this research opportunity, and to the Simons Foundation for funding.

# APPENDIX A

Reflectivity data in the visible and NIR spectrum by target alloy, surface, and coating

| Target ID | Alloy | Surface | Coating | Reflectivity Visible* (counts +/- 10%) | Reflectivity NIR Laser (counts +/- 10%) |
|---|---|---|---|---:|---:|
| 2B6 | 2024 | Beta Cloth | No coating | 29,853 | 29.1 |
| 2B2 | 2024 | Beta Cloth | NIRT Coating | 14,702 | 30 |
| 2G6 | 2024 | Black MLI | No coating | 576,886 | 26.2 |
| 2G2 | 2024 | Black MLI | NIRT Coating | 1,414,759 | 28.4 |
| 2H6 | 2024 | Black MLI and Beta | No coating | 23,477 | 28.1 |
| 2H2 | 2024 | Black MLI and Beta | NIRT Coating | 21,543 | 34.4 |
| 2C6 | 2024 | Copper MLI | No coating | 17,736,121 | 299 |
| 2C2 | 2024 | Copper MLI | NIRT Coating | 3,085,934 | 362 |
| 2D6 | 2024 | Copper MLI and Beta | No coating | 37,250 | 35.1 |
| 2D2 | 2024 | Copper MLI and Beta | NIRT Coating | 18,625 | 40.2 |
| 2A6 | 2024 | Bare Aluminum | No coating | 3,056,840 | 158 |
| 2A2 | 2024 | Bare Aluminum | NIRT Coating | 1,435,044 | 181 |
| 2E6 | 2024 | Silver MLI | No coating | 39,092,905 | 361 |
| 2E2 | 2024 | Silver MLI | NIRT Coating | 6,945,832 | 366 |
| 2F6 | 2024 | Silver MLI and Beta | No coating | 32,414 | 33.3 |
| 2F2 | 2024 | Silver MLI and Beta | NIRT Coating | 24,013 | 40.2 |
| 6B6 | 6061 | Beta Cloth | No coating | 30,509 | 31.4 |
| 6B2 | 6061 | Beta Cloth | NIRT Coating | 18,126 | 29.1 |
| 6G6 | 6061 | Black MLI | No coating | 288,929 | 26.2 |
| 6G2 | 6061 | Black MLI | NIRT Coating | 2,223,779 | 26.7 |
| 6H6 | 6061 | Black MLI and Beta | No coating | 12,678 | 30.2 |
| 6H2 | 6061 | Black MLI and Beta | NIRT Coating | 13,039 | 27.6 |
| 6C6 | 6061 | Copper MLI | No coating | 18,629,731 | 336 |
| 6C2 | 6061 | Copper MLI | NIRT Coating | 3,431,220 | 359 |
| 6D6 | 6061 | Copper MLI and Beta | No coating | 28,475 | 32.1 |
| 6D2 | 6061 | Copper MLI and Beta | NIRT Coating | 15,042 | 41.5 |
| 6A6 | 6061 | Bare Aluminum | No coating | 1,360,031 | 165 |
| 6A2 | 6061 | Bare Aluminum | NIRT Coating | 1,171,692 | 136 |
| 6F6 | 6061 | Silver and Beta | No coating | 21,262 | 36.3 |
| 6F2 | 6061 | Silver and Beta | NIRT Coating | 11,529 | 33.6 |
| 6E6 | 6061 | Silver MLI | No coating | 32,315,529 | 356 |
| 6E2 | 6061 | Silver MLI | NIRT Coating | 3,626,613 | 363 |
| 7B6 | 7075 | Beta Cloth | No coating | 24,882 | 30.8 |
| 7B2 | 7075 | Beta Cloth | NIRT Coating | 21,009 | 38.3 |
| 7G6 | 7075 | Black MLI | No coating | 942,170 | 31.8 |
| 7G2 | 7075 | Black MLI | NIRT Coating | 2,030,523 | 31.3 |
| 7H6 | 7075 | Black MLI and Beta | No coating | 33,880 | 29.2 |
| 7H2 | 7075 | Black MLI and Beta | NIRT Coating | 24,709 | 36.4 |
| 7C6 | 7075 | Copper MLI | No coating | 26,427,562 | 294 |
| 7C2 | 7075 | Copper MLI | NIRT Coating | 4,461,091 | 345 |
| 7D6 | 7075 | Copper MLI and Beta | No coating | 32,344 | 34.4 |
| 7D2 | 7075 | Copper MLI and Beta | NIRT Coating | 24,282 | 34.1 |
| 7A6 | 7075 | Bare Aluminum | No coating | 1,313,416 | 125 |
| 7A2 | 7075 | Bare Aluminum | NIRT Coating | 1,044,208 | 129 |
| 7F6 | 7075 | Silver MLI and Beta | No coating | 34,787 | 33.2 |
| 7F2 | 7075 | Silver MLI and Beta | NIRT Coating | 24,367 | 37.2 |
| 7E6 | 7075 | Silver MLI | No coating | 41,089,052 | 349 |
| 7E2 | 7075 | Silver MLI | NIRT Coating | 7,787,974 | 360 |

* The sum of all of spectroscopy intensity measurements across all wavelengths in the visible light spectrum (380-740 nm), i.e., area under the curve in Appendix B

**APPENDIX B**

All spectrographic data collected, showing reflectivity of halogen light source by Target ID (i.e., plot of data in Appendix C)

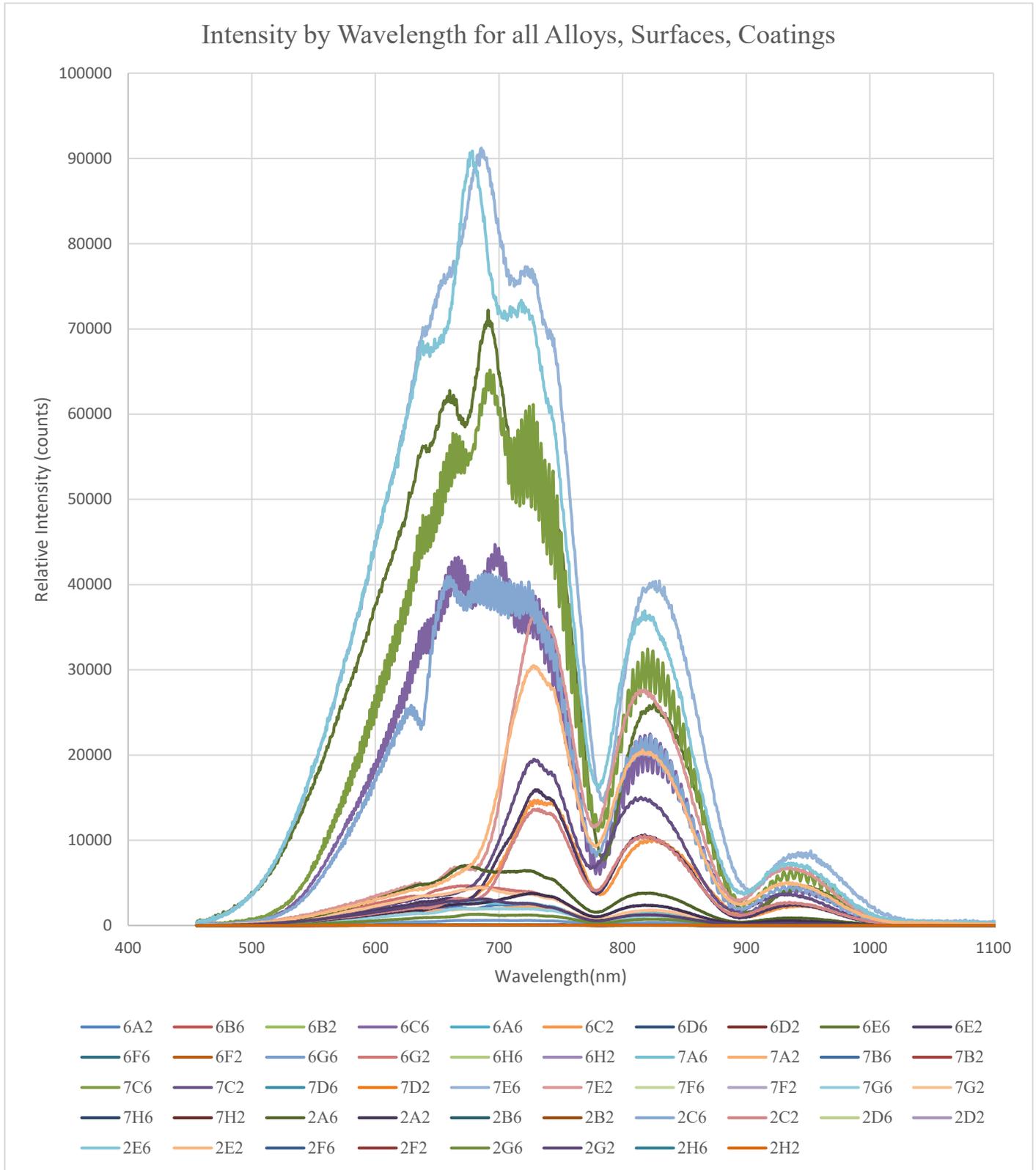

**APPENDIX C**

Excerpt of raw spectroscopy data by Target ID and wavelength

| Wavelength [nm] | 6A5 | 6A4 | 6A3 | 6A2 | 6B6 | 6B5 | 6B4 | 6B3 |
|---:|---:|---:|---:|---:|---:|---:|---:|---:|
| 455.4237 | 24.43191 | 0.7088054 | 1.295392 | 15.18369 | 0.293667 | 45.26474 | 0.869848 | 0.70351 |
| 455.759 | 38.45109 | 0.8771882 | 1.698242 | 16.44641 | 0.378698 | 42.9215 | 1.057995 | 0.943426 |
| 456.0942 | 38.77706 | 0.9741424 | 1.788786 | 15.45782 | 0.597983 | 46.00772 | 1.592157 | 1.245526 |
| 456.4295 | 36.54813 | 0.7907988 | 1.42668 | 10.21884 | 0.644877 | 53.21702 | 1.473103 | 0.882759 |
| 456.7647 | 31.27806 | 0.5909182 | 1.135378 | 9.540333 | 0.63583 | 50.86672 | 1.285241 | 0.441962 |
| 457.0999 | 34.35027 | 0.5409836 | 1.157879 | 10.26646 | 0.520227 | 46.8643 | 1.00211 | 0.505948 |
| 457.4351 | 26.90615 | 0.6766291 | 1.493233 | 12.37951 | 0.602878 | 49.34604 | 1.183973 | 0.912828 |
| 457.7703 | 23.94616 | 0.8806139 | 1.603378 | 13.53037 | 0.630914 | 51.2125 | 1.104549 | 0.851692 |
| 458.1055 | 27.91508 | 1.024785 | 2.214988 | 14.81814 | 0.608285 | 62.66716 | 1.395124 | 1.017536 |
| 458.4407 | 36.2917 | 1.072556 | 1.726704 | 15.17025 | 0.574218 | 61.70879 | 1.462593 | 1.278023 |
| 458.7758 | 42.2628 | 1.020991 | 1.428597 | 11.04424 | 0.552531 | 64.42105 | 1.436339 | 1.117615 |
| 459.111 | 42.91653 | 0.965652 | 1.31157 | 8.27896 | 0.549781 | 58.93748 | 1.310719 | 0.957666 |
| 459.4461 | 40.1829 | 0.8950877 | 1.951195 | 9.112822 | 0.479248 | 58.79679 | 1.197261 | 1.093757 |
| 459.7813 | 30.00328 | 1.022948 | 1.802581 | 15.26094 | 0.546946 | 54.76455 | 0.973648 | 1.132137 |
| 460.1164 | 29.38096 | 1.316057 | 2.544633 | 17.66866 | 0.624403 | 56.15057 | 1.072612 | 1.112431 |
| 460.4515 | 41.05342 | 1.122412 | 2.859687 | 15.54402 | 0.758884 | 61.62801 | 1.199476 | 1.432428 |
| 460.7866 | 52.22399 | 0.8117304 | 2.620995 | 9.211205 | 0.670728 | 54.4374 | 1.120174 | 1.566537 |
| 461.1217 | 45.4036 | 0.5035565 | 1.667276 | 10.82646 | 0.662649 | 48.81047 | 0.797993 | 1.531266 |
| 461.4568 | 37.69089 | 0.6875502 | 1.544369 | 11.03348 | 0.517674 | 53.21233 | 0.919873 | 1.437836 |
| 461.7919 | 46.7975 | 0.7922896 | 1.393397 | 11.75092 | 0.565343 | 65.90971 | 1.064165 | 1.689048 |
| 462.1269 | 57.29439 | 1.222096 | 1.624416 | 11.04457 | 0.514818 | 63.8357 | 1.673007 | 1.644223 |
| 462.462 | 53.72205 | 1.238309 | 1.849859 | 14.40934 | 0.683331 | 62.54272 | 1.698967 | 1.771254 |
| 462.797 | 42.559 | 1.163182 | 1.757502 | 16.68903 | 0.679035 | 60.44577 | 1.57544 | 1.552019 |
| 463.132 | 35.14202 | 1.009524 | 1.789207 | 18.96703 | 0.634288 | 61.17588 | 1.366521 | 1.429006 |
| 463.467 | 32.46576 | 0.9641108 | 2.297479 | 20.95404 | 0.466922 | 53.93919 | 1.748685 | 1.181229 |
| 463.802 | 35.09056 | 0.8569659 | 2.870266 | 18.98986 | 0.619049 | 52.22865 | 1.983943 | 1.378772 |
| 464.137 | 48.27877 | 0.983799 | 2.646597 | 19.24737 | 0.70799 | 51.43147 | 1.808199 | 1.569306 |
| 464.472 | 43.86002 | 1.216559 | 2.217352 | 17.39262 | 0.754165 | 63.94369 | 1.630025 | 1.698481 |
| 464.807 | 40.96027 | 1.294499 | 1.621642 | 18.8378 | 0.61562 | 75.00819 | 1.680417 | 1.771073 |
| 465.1419 | 39.10401 | 1.369197 | 2.284205 | 19.8205 | 0.689393 | 78.03274 | 1.987753 | 1.676447 |
| 465.4769 | 46.00275 | 1.422005 | 2.725974 | 21.93592 | 0.775853 | 85.27383 | 2.276643 | 1.494268 |
| 465.8118 | 50.67449 | 1.233144 | 2.922249 | 20.24355 | 1.065388 | 94.6451 | 2.48964 | 1.607537 |
| 466.1468 | 62.33547 | 1.072676 | 2.571867 | 19.54756 | 0.932626 | 98.30439 | 2.376729 | 1.943979 |
| 466.4817 | 60.64076 | 1.261149 | 2.577951 | 18.8035 | 0.754797 | 87.37145 | 2.133877 | 1.940415 |
| 466.8166 | 46.12587 | 1.433714 | 2.676371 | 18.75482 | 0.573373 | 81.38124 | 1.798208 | 1.759222 |
| 467.1515 | 51.67142 | 1.289204 | 2.592733 | 23.66294 | 0.620673 | 71.78398 | 2.003862 | 1.662133 |
| 467.4864 | 65.12506 | 1.297431 | 2.763827 | 26.57886 | 0.697787 | 77.41381 | 2.092541 | 1.644307 |
| 467.8212 | 71.88157 | 1.487537 | 2.92367 | 28.26748 | 0.784324 | 83.65788 | 2.574664 | 1.628594 |
| 468.1561 | 69.14738 | 2.030953 | 3.031747 | 23.23209 | 0.871885 | 89.48549 | 2.747696 | 2.048259 |
| 468.491 | 62.29805 | 1.951522 | 3.236957 | 20.18131 | 0.818852 | 87.92142 | 2.792562 | 2.306998 |
| 468.8258 | 51.11655 | 1.762372 | 3.393681 | 19.98597 | 0.911756 | 90.46864 | 2.74966 | 2.224126 |
| 469.1606 | 48.83425 | 1.464759 | 3.557402 | 24.80366 | 0.974163 | 86.61021 | 2.689323 | 1.78985 |
| 469.4955 | 58.67112 | 1.611982 | 3.224213 | 25.77099 | 1.001863 | 83.11794 | 2.444162 | 1.779809 |
| 469.8303 | 62.66143 | 1.660961 | 3.440177 | 22.54605 | 0.876022 | 86.61951 | 2.284805 | 1.943035 |
| 470.165 | 66.67404 | 1.806182 | 3.744692 | 22.86512 | 0.835772 | 103.361 | 2.878381 | 2.261679 |
| 470.4998 | 70.77325 | 1.78188 | 3.642293 | 25.70844 | 0.946359 | 108.3386 | 3.235722 | 2.143737 |
| 470.8346 | 67.58793 | 1.684353 | 3.356788 | 30.98186 | 0.932347 | 107.3253 | 3.157205 | 1.765813 |
| 471.1694 | 59.53538 | 1.882852 | 3.349536 | 33.90688 | 0.870935 | 105.0392 | 2.924152 | 2.058826 |
| 471.5041 | 63.47112 | 2.003891 | 3.650298 | 34.19948 | 0.833294 | 113.4388 | 3.018416 | 2.562449 |
| 471.8389 | 68.31236 | 1.831316 | 3.294011 | 30.21088 | 0.97313 | 125.0122 | 3.073472 | 2.664868 |
| 472.1736 | 69.62279 | 1.769666 | 3.593033 | 28.75057 | 1.116656 | 133.4703 | 3.013943 | 2.209756 |
| 472.5083 | 69.15627 | 1.843247 | 3.590756 | 31.42145 | 1.094746 | 132.6649 | 2.760432 | 2.223688 |
| 472.843 | 70.83942 | 1.901176 | 3.941761 | 34.35617 | 0.996557 | 110.5964 | 2.705983 | 2.574879 |
| 473.1777 | 69.50346 | 1.675617 | 3.87642 | 32.30231 | 0.966345 | 108.3912 | 3.019769 | 2.816578 |
| 473.5124 | 77.55138 | 1.685314 | 4.263977 | 33.34013 | 1.048328 | 108.8951 | 3.21948 | 2.80442 |